\documentclass[fleqn,10pt,twocolumn]{wlscirep}
\usepackage[utf8]{inputenc}
\usepackage{color}
\usepackage{babel}
\usepackage{amsmath}
\usepackage{amssymb}
\usepackage{graphicx}
\usepackage{wasysym}
\usepackage{hyperref}
\usepackage{lettrine}
\usepackage{layouts}

\newcommand{\onlinecite}[1]{\hspace{-1 ex} \nocite{#1}\citenum{#1}} 

\newcommand{\GamG}{\Gamma} 
\newcommand{\GamO}{\tilde{\Gamma}} 
\newcommand{\DP}{\Delta P} 
\newcommand{\DPdot}{\Delta \dot{P}} 

\title{Nanobolometer with Ultralow Noise Equivalent Power}
\author[1,*]{R. Kokkoniemi}
\author[1]{J. Govenius}
\author[1,2]{V. Vesterinen}
\author[1,3]{R. E. Lake}
\author[1]{A. M. Gunyho}
\author[1]{K. Y. Tan} %
\author[2]{S. Simbierowicz}
\author[2]{L. Grönberg}
\author[2]{J. Lehtinen}
\author[2]{M. Prunnila}
\author[2]{J. Hassel}
\author[4]{O.-P. Saira} 
\author[1]{M. Möttönen}
\affil[1]{QCD Labs, QTF Centre of Excellence, Department of Applied Physics,
Aalto University, P.O. Box 13500, FIN-00076 Aalto, Finland}
\affil[2]{VTT Technical Research Centre of Finland Ltd. \& QTF Centre of Excellence, P.O. Box 1000, 02044 VTT, Finland}
\affil[3]{National Institute of Standards and Technology, Boulder, Colorado, 80305, USA}
\affil[4]{Department of Applied Physics, California Institute of Technology, MC 149-33, Pasadena, California, 91125, USA}
\affil[*]{roope.kokkoniemi@aalto.fi}

\date{\today}

\begin{abstract}
{\bf Since the introduction of bolometers more than a century ago, they have been applied in a broad spectrum of contexts ranging from security and the construction industry to particle physics and astronomy. However, emerging technologies and missions call for faster bolometers with lower noise. Here, we demonstrate a nanobolometer that exhibits roughly an order of magnitude lower noise equivalent power, $20\textrm{ zW}/\sqrt{\textrm{Hz}}$, than previously reported for any bolometer. Importantly, it is more than an order of magnitude faster than other low-noise bolometers, with a time constant of 30~$\mu$s at $60\textrm{ zW}/\sqrt{\textrm{Hz}}$. These results suggest a calorimetric energy resolution of $0.3\textrm{ zJ}=h\times 0.4$~THz with a time constant of 30 $\mu$s. 
Thus the introduced nanobolometer is a promising candidate for future applications requiring extreme precision and speed such as those in astronomy and terahertz photon counting.}
\end{abstract}
\begin{document}

\flushbottom
\maketitle
\lettrine[lines=3]{\textcolor{red}{M}}{}easuring electromagnetic radiation at different frequencies and polarizations
is one of the few experimental tools available for studying the universe
at extragalactic scales. Consequently, improved detector sensitivity
has been a key factor in enabling major cosmological breakthroughs,
such as the observation of minuscule anisotropies in the cosmic microwave
background~\cite{lamarre2010planck}. Especially at microwave and far-infrared frequencies,
moving the observatories to space has been pivotal in enabling these
observations. As a next step, space-based observatories will have
larger and colder telescopes. 
Proposed and planned future missions call for~\cite{Jackson2011SPICASAFARI,Karasik2011Nanobolometers,baselmans2017kilo} noise equivalent power (NEP) of the order of $10\mbox{ zW/\ensuremath{\sqrt{\mbox{Hz}}}}$ or lower.

Among different techniques for measuring the power carried by an electromagnetic signal, we focus on bolometers,
i.e., devices that detect radiation-generated heat in
an absorber. Bolometry is one of the oldest radiation sensing techniques~\cite{langley1880bolometer} dating back to 1880 and yet remains competitive and widespread~\cite{armengaud2016lumineu}, mainly owing to the flexibility
bolometers offer in terms of the center frequency, bandwidth, and dynamic
range, as well as the possibility of energy-resolving calorimetric
operation~\cite{gray2016first}. Furthermore, thermal conductances below $1\mbox{ fW/K}$
between modern nanoscale bolometers and their environment have been
measured~\cite{Wei2008Ultrasensitive,Govenius2016Detection}, implying
that the NEP limit set by thermal energy fluctuations can be reduced
down to at least $10\mbox{ zW/\ensuremath{\sqrt{\mbox{Hz}}}}$.
To date, however, the lowest measured NEPs for bolometers are
around $300\mbox{ zW/\ensuremath{\sqrt{\mbox{Hz}}}}$~\cite{Karasik2011Demonstration,Suzuki2014Performance}, achieved using transition edge sensors~\cite{Ullom2015Review} (TESs).

Promising NEPs have also been reported for kinetic inductance
detectors ($400\mbox{ zW/\ensuremath{\sqrt{\mbox{Hz}}}}$)~\cite{deVisser2014Fluctuations}
and proof-of-principle quantum capacitance detectors (of the order of $10\mbox{ zW/\ensuremath{\sqrt{\mbox{Hz}}}}$)~\cite{Echternach2013Photon}.
Both of these detect radiation-generated non-equilibrium quasiparticles in a superconductor.
In addition, the background rate of the random telegraph noise in
semiconducting charge sensors shows potential for extremely low NEP
($\apprle 1\mbox{ zW/\ensuremath{\sqrt{\mbox{Hz}}}}$)~\cite{Komiyama2000Singlephoton,Komiyama2011SinglePhoton}.
However, the coupling efficiency to a radiation source is expected
to be low and experimental characterization of the efficiency has
not been reported. 

In this work, we implement a continuously operating bolometer based on superconductor--normal-metal--superconductor (SNS) junctions (Fig.~\ref{fig:setup}a,~b), and measure an $\mbox{NEP}\approx50\mbox{ zW/\ensuremath{\sqrt{\mbox{Hz}}}}$ with a time constant of $0.6\mbox{ ms}$ (Fig.~\ref{fig:resp}g). We can also tune the time constant, in situ, below $100~\mu \mathrm{s}$ at the expense of increasing
the NEP to $80\mbox{ zW/\ensuremath{\sqrt{\mbox{Hz}}}}$,
which is nevertheless still lower than the lowest previously reported
bolometer NEPs~\cite{Karasik2011Demonstration,Suzuki2014Performance}.
By introducing a Josephson parametric amplifier~\cite{vesterinen2017lumped} (JPA) to the bolometer readout circuit (Fig.~\ref{fig:setup}a,~c), we further reduce the NEP to a record low value of $20\mbox{ zW/\ensuremath{\sqrt{\mbox{Hz}}}}$ (Fig.~\ref{fig:resp}h).
For an NEP of $60\mbox{ zW/\ensuremath{\sqrt{\mbox{Hz}}}}$, we achieve response times down to about 30~$\mu$s using the JPA, which is one to two orders of magnitude faster than those reported for the most sensitive TESs~\cite{Suzuki2014Performance}. Although the experiments described here focus on the measurement of the input power, the achieved NEP and time constant suggest that in a calorimetric mode of operation, the energy resolution can be well below the current 1.1-zJ record for thermal detectors~\cite{Govenius2016Detection} (Fig.~\ref{fig:noise}c).
\subsection*{Device and measurement setup}
\begin{figure*}[htb!]
\includegraphics[width=\textwidth]{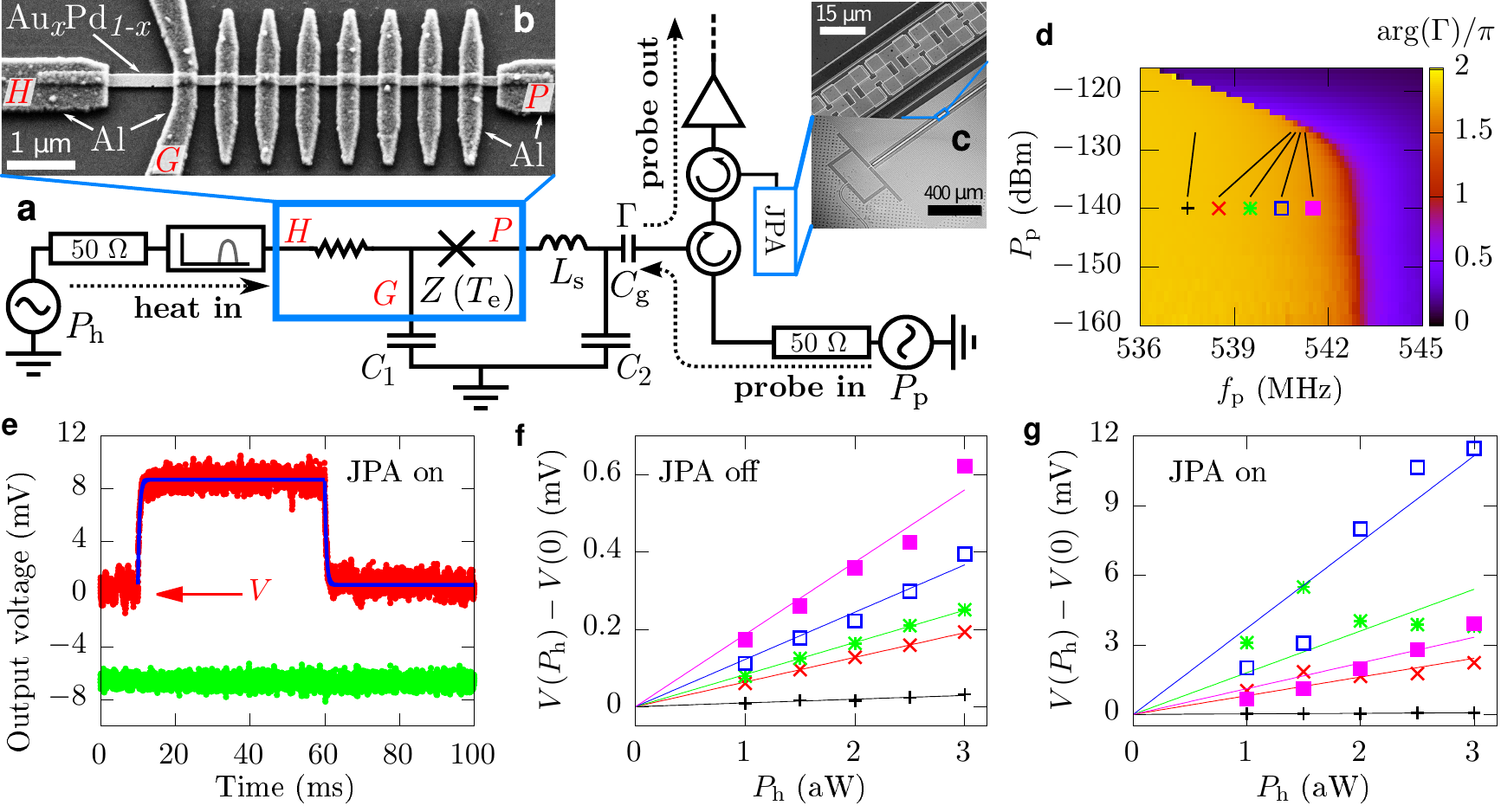}

\caption{{\bf Measurement setup and device characterization.} \textbf{a--c}, Simplified detector measurement setup (\textbf{a}) together with micrographs of the SNS junctions between leads $H$, $G$, and $P$ taken from a similar device (\textbf{b}) and of a similar JPA (\textbf{c}). The impedance
$Z(T_\mathrm{e})$ of the series of short SNS junctions forms a temperature-sensitive
resonant circuit together with a meander inductor $L_\mathrm{s}\approx1.2\mbox{ nH}$
and the capacitors $C_{1}\approx87\mbox{ pF}$ and $C_{2}\approx33\mbox{ pF}$.
The gate capacitance is $C_\mathrm{g}\approx0.87\mbox{ pF}$. There is an 8.4-GHz bandpass filter connected to lead H. \textbf{d}, Phase of the reflection coefficient at the gate capacitor $\GamG$ as a function of probe frequency $f_\mathrm{p}$ and power $P_\mathrm{p}$, without heating and with the JPA off.  \textbf{e}, Example of the ensemble-averaged detector output voltage at the digitizer $V$ (red colour) and voltage in the other quadrature (green colour)
, with the JPA on. 
The blue curves show exponential fits to the rising and falling edges of the signal.  \textbf{f,~g}, Change in the detector output voltage after the heater is turned on (see panel \textbf{e})  
as a function of the finite heater power $P_\mathrm{h}$ with the JPA off~(\textbf{f}) and on~(\textbf{g}). The $(f_\mathrm{p},P_\mathrm{p})$ operation points are indicated in panel~\textbf{d}. The bath temperature is $T_\mathrm{b}=25\mbox{ mK}$ for all data in this paper.}\label{fig:setup}
\end{figure*}

Figure~\ref{fig:setup} shows the detector and its measurement scheme.  
We couple the detector to an $8.4\mbox{-GHz}$ microwave
source through a $50\,$-$\Omega$ transmission line, which allows us
to calibrate the heater power $P_\mathrm{h}$ incident at the detector input with a decibel level of uncertainty. 
Essentially all heater power is absorbed by the long SNS junction between leads $H$ and $G$ since the junction is long enough for its impedance to be essentially real, $36\mbox{ \ensuremath{\Omega}}$, and well matched to the transmission line impedance of $Z_0 = 50\ \Omega$. Thus an increase in $P_{\mathrm{h}}$ leads to an increase in the electron temperature $T_\mathrm{e}$ in the $\textnormal{Au}_{x}\textnormal{Pd}_{1-x}$ nanowire used as the normal-metal part in the SNS junctions. This in turn results in an increased inductance of the short SNS junctions~\cite{Giazotto2008Ultrasensitive} between leads $P$ and $G$, which implies a lower resonance frequency of the effective $LC$ oscillator formed by the short SNS junctions, the on-chip meander inductor $L_\mathrm{s}$, and the on-chip parallel plate capacitors $C_{1}$ and $C_{2}$. We detect this change by measuring the reflection coefficient at the detector gate capacitor $\GamG\left(T_\mathrm{e},\omega_\mathrm{p}\right)$ 
at a fixed probe frequency $f_\mathrm{p}=\omega_\mathrm{p}/2\pi$. See Methods for the extraction of $\GamG$ from the detector output voltage at the digitizer $V$.
Furthermore, we have the option to amplify the readout signal with a JPA [Fig.~\ref{fig:setup}(c), ref.~\onlinecite{simbierowicz2018flux}, see Methods].

\subsection*{Characterization experiments}
Figure~\ref{fig:setup}d shows the phase of the reflection coefficient at the gate capacitor (see Methods for details) as a function of probe frequency and probe power, at zero heater power. The most striking feature in Fig.~\ref{fig:setup}d is the decreasing resonance frequency as the probe tone begins to significantly heat the electrons in the SNS junctions above $P_\textrm{p}=-135$~dBm. This electrothermal feedback~\cite{deVisser2010Readoutpower} can be used to optimize the sensitivity and the time constant of the detector or even induce temperature bistability~\cite{Govenius2016Detection} (not visible in Figure~\ref{fig:setup}d).

The NEP is determined by how noisy the readout signal is relative to the responsivity of the signal to changes in $P_\mathrm{h}$ (see Methods). Thus in Fig.~\ref{fig:setup}e, we show an example of the detector output voltage $V$ which is defined as the voltage in the quadrature providing the greatest response to the heater power after amplification ($\approx103\mbox{ dB}$), demodulation, and an optimally chosen phase rotation.
In Fig.~\ref{fig:setup}e, we first set 
$P_\mathrm{h}$ to  a small value ($3\mbox{ aW}$) for a period of several tens of ms, then turn $P_\mathrm{h}$ off for a similar period, and finally average over repetitions of this modulation pattern. From such data we extract the quasistatic voltage response at the digitizer and 
the time constant using exponential fitting functions. 

Figure~\ref{fig:setup}f,~g shows the quasistatic response of the detector output voltage to the heater power up to 3 aW. We define the detector responsivity as the ratio of the voltage response and the corresponding heater power. 
We employ this information to choose an appropriate power level for the heater in our experiments discussed below. 
\subsection*{Dimensionless susceptibility}
To understand the detector response at high probe power, we develop a model for the electrothermal feedback\cite{Govenius2016Detection,Govenius2016Towards}. (See Methods for details.) We define a dimensionless susceptibility as
\begin{equation}
\chi=\left.\frac{\partial\DP}{\partial P_{\textnormal{h}}}\right|_{\partial_t \DP=0},\label{eq:chi-def}
\end{equation}
where 
$\DP$ equals the amount of additional heat flowing from the nanowire electrons to their thermal bath. Therefore, $\chi$ is the factor by which the probe-induced electrothermal feedback enhances the heating of the bolometer relative to the externally applied power $P_\mathrm{h}$. 

\subsection*{Detector responsivity and noise}
In Fig.\ref{fig:resp}a,~b, we show the responsivity of the detector output voltage. Note that the NEP is unaffected by the calibration of the gain of the readout circuitry since the measured responsivity and noise are both amplified equally.
The responsivity is maximized for probe frequencies close to the resonance. 
As the probe power is increased, the width of the resonance decreases, leading to a sharp increase in the responsivity. Note that the color scales are different in Fig.~\ref{fig:resp}a,~b, since the JPA adds gain in excess of 20 dB.

Interleaved with these measurements of the responsivity, we also record separate noise spectra for the detector output voltage and the out-of-phase quadrature at each probe power and frequency. In Fig.~\ref{fig:resp}c,~d, we show the voltage noise spectral density across the same range of $f_\mathrm{p}$ and $P_\mathrm{p}$. Let us first discuss the low-probe-power limit ($P_\mathrm{p}\apprle-132\mbox{ dBm}$) with the JPA off in Fig.~\ref{fig:resp}c. Here, the electrothermal feedback is negligible ($\chi\approx1$), $\DP$ vanishes, and the spectrum is dominated by noise added by the amplifiers in the readout circuitry. In this case, the noise power assumes a similar value on and off resonance. However, with the JPA on in Fig.~\ref{fig:resp}d, we consistently observe a peak in the noise near the resonant probe frequency, indicating that amplifier noise is not dominating the signal even at the lowest probe powers shown ($-132.5$~dB). With the JPA off, the thermal fluctuations of the detector surpass the amplifier noise only at high $P_\mathrm{p}$.

\subsection*{Noise equivalent power and time constant}
\begin{figure*}
\includegraphics[width=\textwidth]{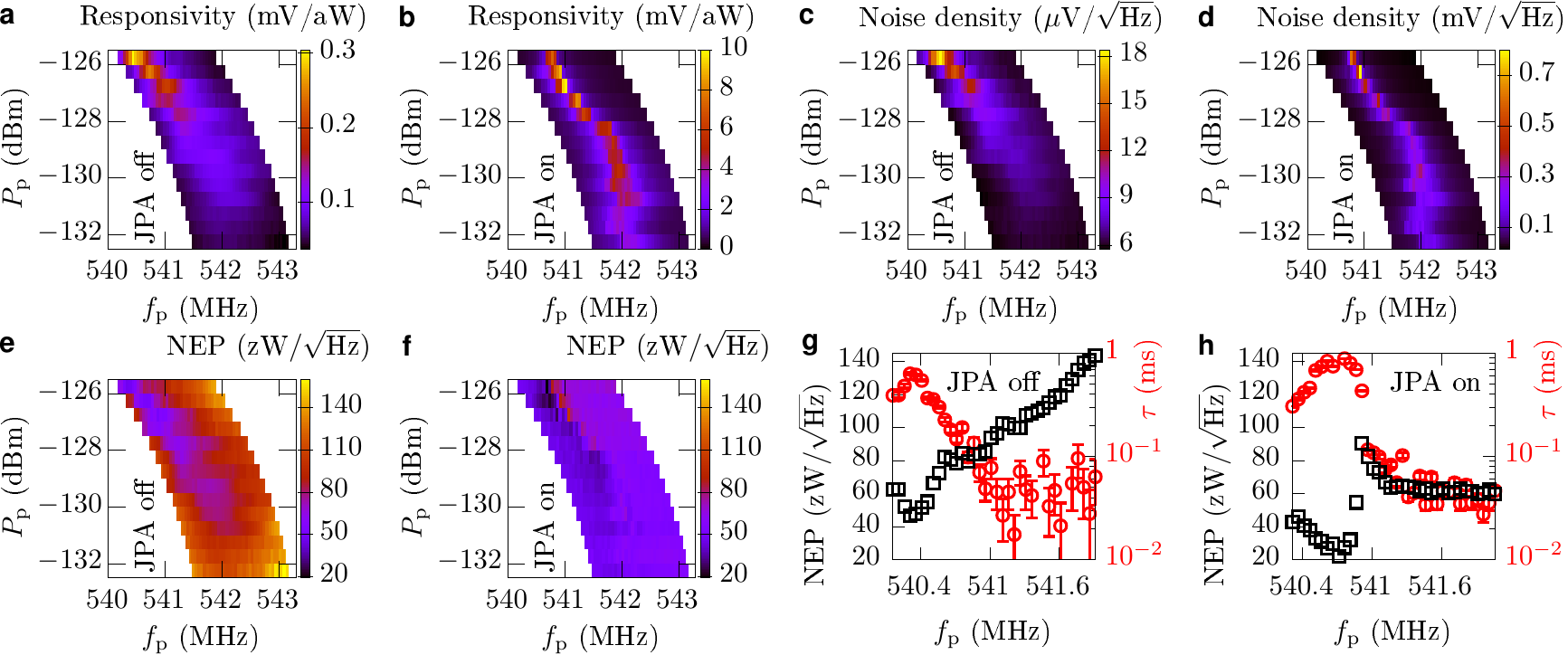}

\caption{{\bf Noise equivalent power and thermal time constant.} \textbf{a--d}, Quasistatic responsivity of the probe voltage to the heater power (\textbf{a,~b}) and probe voltage spectral density (\textbf{c,~d}) as functions of probe frequency and power with the JPA off (\textbf{a,~c}) and on (\textbf{b,~d}) averaged over noise frequencies between 20 and 100 Hz. \textbf{e,~f}, Noise equivalent power (NEP) as a function of probe frequency and power with the JPA off (\textbf{e}) and on (\textbf{f}) averaged over noise frequencies between 20 and 100~Hz. \textbf{g,~h}, NEP (black) and thermal time constant (red) with the JPA off at fixed $P_\mathrm{p}=-126\text{ dBm}$ (\textbf{g}) and with the JPA on at $P_\mathrm{p}=-126.5\text{ dBm}$ (\textbf{h}).
\label{fig:resp}}
\end{figure*}

Figure~\ref{fig:resp}e,~f presents the main results of this paper, i.e., the measured NEP for an 8.4-GHz input 
over a range of probe powers and frequencies. We compute the NEP by dividing the voltage spectral density by the quasistatic responsivity and multiplying the result by a factor $\sqrt{1+(2\pi\tau f_\mathrm{n})^2}$, where $f_\mathrm{n}$ is the noise frequency. This factor takes into account the fact that the thermal time constant $\tau$ decreases the responsivity of the detector with respect to the quasistatic case (see Methods). Figure~\ref{fig:resp}e,~f shows the NEP with the JPA on and off, respectively, averaged over noise frequencies from 20 Hz to 100 Hz. 

In Fig.~\ref{fig:resp}g we show the NEP and the time constant as functions of probe frequency at fixed $P_\mathrm{p}$ of $-126$ dBm with the JPA off. Figure~\ref{fig:resp}h is measured in identical conditions except that the JPA is on and the probe power set to $-126.5$ dBm. The electrothermal feedback is strong and positive ($\chi\gg 1$) at probe frequencies just below the resonance frequency. By contrast, the electrothermal feedback is strongly negative ($\chi\ll 1$) at probe frequencies just above the resonance. This is clearly visible in the time constant $\tau=\chi\tau_\textrm{b}$ 
which increases by nearly an order of magnitude as the probe approaches the resonance 
despite the fact that the bare thermal time constant $\tau_\textrm{b}$ simultaneously decreases owing to increased electron temperature. Here, $\tau_\textrm{b}$ denotes the time constant in the absence of electrothermal feedback 
(see Methods).
The lowest NEP of $20 \mbox{ zW/\ensuremath{\sqrt{\mbox{Hz}}}}$ in Fig.~\ref{fig:resp}h coincides with the peak of the time constant ($1\mbox{ ms}$), suggesting that at $P_\mathrm{p}=-126.5\mbox{ dBm}$ the NEP is optimized at the frequency that maximizes $\chi$.

As the probe frequency exceeds the resonance, the time constant quickly decreases by more than an order of magnitude below $100\mbox{ \ensuremath{\mu}s}$. In this regime, the positive effect of the JPA is particularly clear, as the NEP  degrades quickly with increasing probe frequency when the JPA is disabled, but stays roughly constant when it is enabled. The fact that the NEP remains relatively flat at $60\mbox{ zW/\ensuremath{\sqrt{\mbox{Hz}}}}$ with the JPA on (Fig.\ref{fig:resp}h) is an indication that the internal fluctuations of the detector are limiting the performance instead of amplifier noise. Even here the NEP is below that of other state-of-the-art bolometers~\cite{Karasik2011Demonstration,Suzuki2014Performance}. This is an example of the convenient \emph{in situ} tunability of the SNS detector, i.e., we can choose a different trade off between the NEP and the time constant by a small change of the probe frequency or power. We can also tune the time constant and the dynamic range by changing the bath temperature or by applying an additional constant heating power through the heater port. However, this optimization is left for future work.

\begin{figure}
\includegraphics[width=\columnwidth]{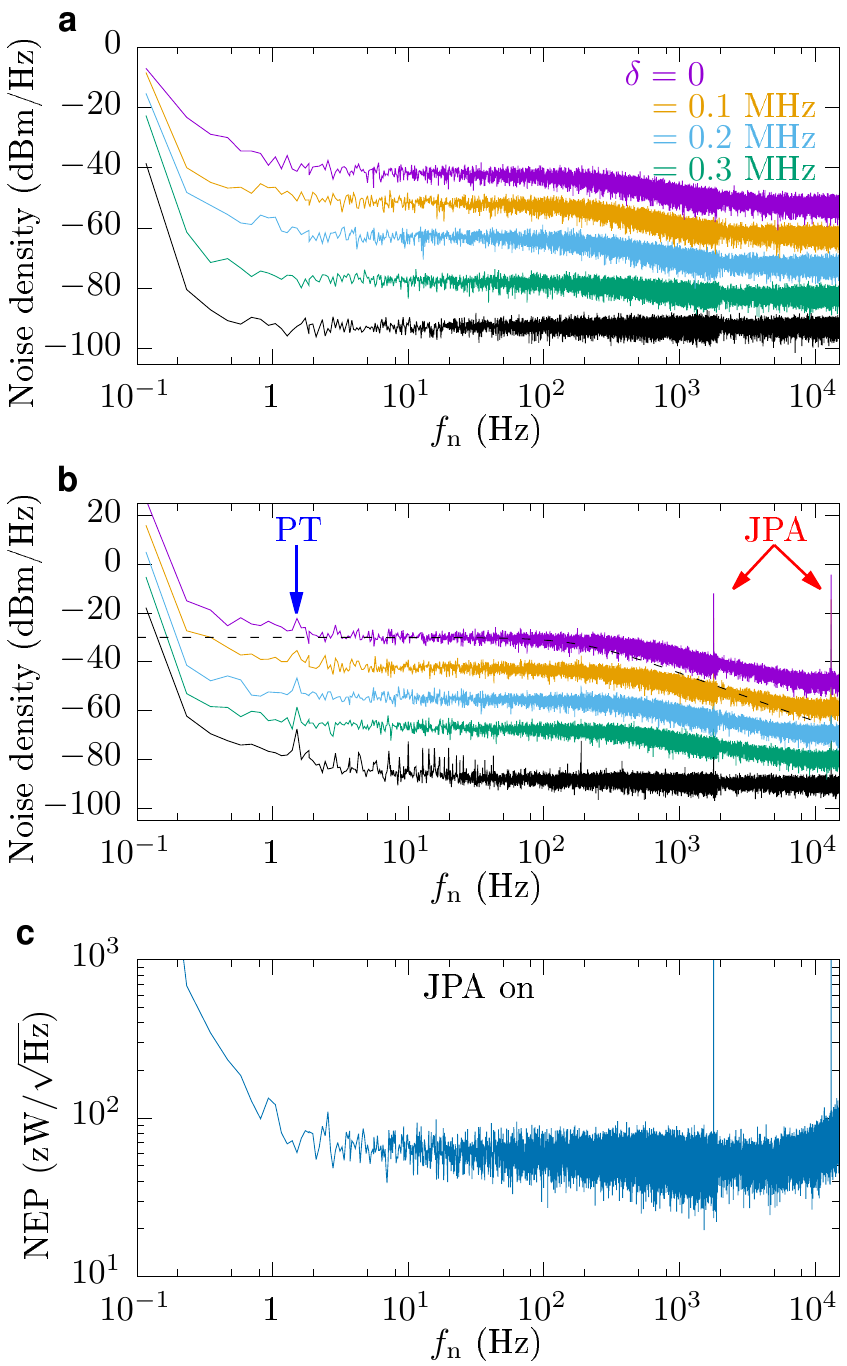}

\caption{{\bf Frequency spectra of voltage noise and NEP.} \textbf{a,~b}, Spectral density of the noise in the signal quadrature of the down-converted probe tone with the JPA off (\textbf{a}) and on (\textbf{b}) as functions of the noise frequency $f_\mathrm{n}$. The bottom most curve shows the spectral density far off resonance at $f_ \mathrm{p}=539.275\mbox{ MHz}$ and $P_\mathrm{p}=-132.5\mbox{ dBm}$, whereas the green, blue, orange, and yellow curves are measured at $P_\mathrm{p}=-126\mbox{ dBm}$ (JPA off) and $P_\mathrm{p}=-126.5\mbox{ dBm}$  (JPA on), and span a narrow frequency range near the resonance. Specifically, the probe frequencies are $f_\mathrm{p}=540.6125 \mbox{ MHz} - \delta$, where the values of $\delta$ are indicated in panel~\textbf{a}. For clarity, the curves have been offset vertically in increments of 10 dBm/Hz. The two peaks above 1 kHz in panel~\textbf{b} are due to the aliased JPA idler. The excess noise at multiples of 1.4 Hz is attributed to the Cryomech pulse tube (PT). The dashed line indicates a first-order RC filter response with a time constant identical to that in the $\delta=0$ trace in panel~\textbf{b}. \textbf{c}, NEP with the JPA on as a function of the noise frequency at $P_\mathrm{p}=-126.5$ dBm and $f_\mathrm{p} = 541.9625$ MHz. These data yield 0.3~zJ for the energy resolution estimate of the detector (see text). 
	Discontinuity in the data on all panels near 2 kHz is caused by the fact that we measure the high and low frequency noise separately with different time steps. }\label{fig:noise}
\end{figure}

\subsection*{Noise analysis}
In Fig.~\ref{fig:noise}, we present the full noise spectrum of the output signal at $P_\mathrm{p}=-126\mbox{ dBm}$ and $P_\mathrm{p}=-126.5\mbox{ dBm}$ with the JPA off and on, respectively. Above 1 Hz and below 1 kHz, the noise increases up to $14.5$ dB above the broadband background set by the amplifier noise for probe frequencies near resonance. Far off-resonance we find only the broadband amplifier noise floor in addition to $1/f_\mathrm{n}$ noise. We also observe in Fig.~\ref{fig:noise}b noise peaks at multiples of $1.4\mbox{ Hz}$, matching the frequency of vibrations caused by the Cryomech pulse tube cooler. Note that these peaks are clearly visible only when the JPA is on and the probe is far from the resonance,  suggesting that the pulse tube noise does not couple directly to the detector, but rather to the amplifiers. At operation points with low NEP, the pulse tube noise is masked by the noise generated by the detector itself.

\subsection*{Predicted energy resolution}
In Fig.~\ref{fig:noise}c, we present the NEP measured with the JPA on as a function of the noise frequency at a $(f_\mathrm{p},P_\mathrm{p})$ point selected for short time constant and low NEP. From the NEP, we can obtain an estimate for an upper bound on the energy resolution~\cite{Moseley1984Thermal}
\[
\epsilon \approx \left(\int_0^\infty\frac{ 4\mathrm{d}f}{\textnormal{NEP}(f)^2}\right)^{-1/2}.
\]
By restricting the above frequency integration below the thermal cut off frequency $1/(2\pi \tau)=5.8\ \mathrm{kHz}$, the data in Fig.~\ref{fig:noise}c yields $\epsilon=0.32\ \mathrm{zJ} = h \times 480\ \mathrm{ GHz}$, surpassing, e.g. the anticipated resolution of the TES-based Fourier transform spectrometer~\cite{Jackson2011SPICASAFARI} specified to have about an octave of resolution in the band of 1.4--9 THz. Increasing the cut off frequency to 10 kHz yields $\epsilon=0.26\ \mathrm{zJ} = h \times 390\  \mathrm{GHz}$.

\subsection*{Conclusions}
We have measured an NEP of $20 \mbox{ zW/\ensuremath{\sqrt{\mbox{Hz}}}}$
for an SNS-junction-based bolometer. This
is an order of magnitude improvement over results reported for transition
edge sensors and kinetic inductance detectors (KIDs). In this sensitivity range, we also report response times down to 30~$\mu$s, i.e., detector bandwidth of about 5 kHz. For comparison, the most sensitive TESs and KIDs operate at bandwidths of order 100 Hz~\cite{Jackson2011SPICASAFARI,deVisser2014Fluctuations}. 
The observed detector speed and NEP predict an energy resolution compatible with single-photon detection extending down to the low THz range, $\sim\! 400$~GHz. 

These results render the SNS detector an attractive candidate for astronomical applications in the microwave and THz regimes. To this end, we aim to accommodate the SNS detector with suitable antennas in the future, which we consider feasible owing to the real detector input impedance. The impedance may be varied and matched by the choice of the normal-metal size and aspect ratios. Furthermore, it is straightforward to frequency multiplex the rf readout of SNS detectors.  

Note that the adaptation of the SNS detector to THz applications may require a higher-gap superconductor than aluminum in the input waveguide or antenna to avoid excess absorption of radiation far away from the SNS junctions. However, this technical change can be carried out such that it has a negligible effect on the SNS junctions~\cite{wang2001terahertz}. 

Recently, superconducting qubits have shown great progress in detecting single microwave photons~\cite{Inomata2016Single,PhysRevX.6.031036,besse2017single,kono2018quantum}, but this technology is currently limited to frequencies below 10 GHz. Thus the SNS detector provides a valuable complementary approach. 

More speculatively, the SNS detector can also be considered a candidate for the light dark matter experiments proposed theoretically in ref.~\onlinecite{Hochberg2016Detecting},
i.e., the detector could be used to detect dark-matter-generated quasiparticles
that diffuse into the SNS junctions after being generated in
the aluminum capacitor plates ($C_{1}$ and $C_{2}$). However, this application calls for a careful estimation of the dark and total counts, accompanied by an optimization of the absorber efficiency.

\section*{Methods}

\subsection*{Sample fabrication and measurement setup}
For details of the sample fabrication methods and of the measurement setup, see ref.~\onlinecite{Govenius2016Detection}. 

\subsection*{Reflection coefficient}
Before feeding the probe signal into the cryostat, we split a fraction of it into a reference tone. We digitize both signals, the reference and the eventual probe signal which is reflected from the detector gate and subsequently amplified and guided out of the cryostat.
We define 
$\GamO\left(T_\mathrm{e},\omega_\mathrm{p}\right)$ as the ratio of the reflected signal and the reference signal. We obtain the reflection coefficient at the gate capacitor $\GamG\left(T_\mathrm{e},\omega_\mathrm{p}\right)$ by first measuring $\GamO\left(T_\mathrm{e},\omega_\mathrm{p}\right)$ with high probe and heating power ($\sim -120$ dBm) and dividing the subsequent measurements by this high-power reference, i.e., $\GamG\left(T_\mathrm{e},\omega_\mathrm{p}\right)=\GamO\left(T_\mathrm{e},\omega_\mathrm{p}\right) / \GamO\left(T \gg T_\mathrm{e},\omega_\mathrm{p}\right)$. The high power shifts the resonance far from its low-power position, thus providing an accurate calibration for the low-power experiments. 

\subsection*{Josephson parametric amplifier}
The utilized JPA is that referred to as Device A in ref.~\onlinecite{simbierowicz2018flux} which provides further details. It is a lumped-element rf resonator where an array of 200 superconducting quantum interference devices (SQUIDs) forms a non-linear inductor. The SQUIDs are dc biased and rf pumped with magnetic flux, generating a three-wave mixing process in the JPA. We use the JPA in the non-degenerate mode where the flux pump is at 2($f_\mathrm{p} + 21.875$ kHz). As the pump is at approximately twice the bolometer resonance frequency, we avoid residual heating of the bolometer by the JPA.

\subsection*{Thermal time constant and electrothermal feedback}
In the low-$P_\mathrm{p}$ limit $\GamG$ is independent of the probe power.
If the probe power is increased, however, the power $(1-|\GamG(T_\mathrm{e},\omega_\mathrm{p})|^{2})P_{\textnormal{p}}$
absorbed from the probe starts to significantly heat the bolometer and shifts
the resonance to a lower frequency, as shown in Fig.~\ref{fig:setup}c. More precisely, the nanowire electron temperature $T_\mathrm{e}$ is determined by 
\begin{align}
C_\mathrm{e}(T_\mathrm{e})\dot{T}_\mathrm{e} & =-P_{\mathrm{e}-\textnormal{b}}(T_\mathrm{e},T_{\textnormal{b}})+P_\mathrm{x}+P_{\textnormal{h}}\label{eq:dTedt}\\
& \qquad+(1-|\GamG(T_\mathrm{e},\omega_\mathrm{p})|^{2})P_{\textnormal{p}},\nonumber 
\end{align}
if we model the electrons in the nanowire using a single heat capacity
$C_\mathrm{e}$ and assume that the electrical degrees of freedom relax to
a quasistationary state quickly compared to the thermal relaxation
time~\cite{Govenius2016Detection}. Here, $P_{\mathrm{e}-\textnormal{b}}(T_\mathrm{e},T_{\textnormal{b}})$
is the heat flow from the electrons to their thermal environment at temperature $T_\mathrm{b}$ and
$P_\mathrm{x}$ is a constant parasitic heating term arising from uncontrolled noise sources.

Let us analyze the increase in the power flow from the nanowire electrons
to their thermal environment, as compared to the case $P_\mathrm{h}=P_\mathrm{p}=0$.
We define this increase as 
\begin{equation}
\DP(T_\mathrm{e})=P_{\mathrm{e}-\textnormal{b}}(T_\mathrm{e},T_{\textnormal{b}})-P_\mathrm{x}.\label{eq:Delta}
\end{equation}
It is convenient to discuss $\DP$ rather than $T_\mathrm{e}$ because
$\DP$ can be directly measured~\cite{Govenius2016Detection}
and it allows us to simplify equation~\eqref{eq:dTedt} to
\begin{equation}
\tau_\textrm{b}\left(\DP\right)\DPdot=-\DP+P_{\textnormal{h}}+(1-\left|\GamG\left(\DP,\omega_\mathrm{p}\right)\right|^{2})P_{\textnormal{p}},\label{eq:dDeltadt}
\end{equation}
where $\tau_\textrm{b}\left(\DP\right)=C\left(\DP\right)\slash\partial_{T_\mathrm{e}}P_{\mathrm{e}-\textnormal{b}}[T_\mathrm{e}(\DP),T_{\textnormal{b}}]$
is the bare thermal time constant~\cite{Govenius2016Detection}, not including the effects of the electrothermal feedback, and $\DPdot$ denotes $\partial_{t}\Delta P$.

We concentrate on the non-bistable regime where equation~\eqref{eq:dDeltadt}
has a unique stationary solution. In this regime, we can define the
single-valued dimensionless susceptibility given by equation~\eqref{eq:chi-def}. The susceptibility also allows us to further simplify equation~\eqref{eq:dDeltadt}
into
\begin{equation}
\chi(\omega_\mathrm{p},P_\mathrm{p})\tau_\textrm{b}(\DP_{0})\partial_{t}(\DP-\DP_{0})\approx-(\DP-\DP_{0}),\label{eq:chi-tau}
\end{equation}
for small deviations around $\DP=\DP_{0}$ that solves equation~\eqref{eq:dDeltadt}
in steady state~\cite{Govenius2016Towards}. From equation~\eqref{eq:chi-tau}
we observe that the effective thermal time constant is given by
$$ 
\tau=\chi(\omega_\mathrm{p},P_\textrm{p})\tau_\textrm{b}(\DP_{0}).
$$

\subsection*{Noise equivalent power}
We define $\textrm{NEP}^2(f_\textrm{n})$ of a noisy bolometer as the one-sided power spectral density of input power fluctuations (units: $\textrm{W}^2/\textrm{Hz}$) that yields for an ideal bolometer a noise spectral density 
in the output signal identical to that of the noisy bolometer. Here, the ideal bolometer refers to a noiseless conversion of input power into output signal with a responsivity equal to that of the noisy bolometer. Equivalently,
$2\sqrt{2\textrm{B}} \times \textrm{NEP}(f_\textrm{n})$ describes, in a narrow bandwidth $\textrm{B}$ centered at $f_\textrm{n}$, the
peak-to-peak amplitude by which the input power needs to be modulated at
$f_\textrm{n}$ for unit signal-to-noise ratio in the output.

Specifically in this work, the noise equivalent power (shown in Fig.~\ref{fig:resp}e--h) is given by
\begin{equation}
\textrm{NEP}(f_\textrm{n}) = \frac{\sqrt{S_V(f_n)}}{R_{P\to V}(f_\textrm{n})}, \label{eq:NEPnew}
\end{equation}
where $R_{P\to V}(f_\textrm{n})$ is the frequency-dependent
responsivity (shown in Fig.~\ref{fig:resp}a,~b) and $S_V(f_\textrm{n})$ is the single-sided power
spectral density of the output signal $V$ (Fig.~\ref{fig:resp}c,~d shows $\sqrt{S_V}$). Note that $\textrm{B}
\times S_V(f_\textrm{n})$ equals the ensemble variance if the signal is filtered
to a narrow bandwidth $\textrm{B}$ centered at $f_\textrm{n}$.

In practice, we measure $S_V(f_\textrm{n})$ by averaging periodograms
according to Bartlett's method and determine the frequency-dependent responsivity from 
\begin{equation}
R_{P\to V}(f_\textrm{n})=\frac{|\partial_{P_\textrm{h}}V|}{\sqrt{1+(2\pi f_\textrm{n}\tau)^2}},\label{eq:resp}
\end{equation}
where $|\partial_{P_\textrm{h}}V|$ is the measured quasistatic responsivity and $\tau$ is the measured time constant. Based on the adequate quality of the fits used to extract
$\tau$ (see Fig.~\ref{fig:setup}e), this single-time-constant
approximation is justified at least up to frequencies of the order of $1/(2
\pi \tau)$. We note that equations~\eqref{eq:resp} and~\eqref{eq:NEPnew} are identical to those used for NEP in previous literature~\cite{deVisser2014Fluctuations, deVisserThesis}.

\subsection*{Data availability}
The data that support the findings of this study are available from the corresponding
author upon reasonable request.
\bibliography{nep}

\section*{Acknowledgments}
We  acknowledge  the  provision  of facilities  and  technical  support  by  Aalto  University  at OtaNano --- Micronova  Nanofabrication  Centre. We  have received  funding  from  the  European  Research  Council  under  Consolidator  Grant No.  681311 (QUESS) and under Proof of Concept Grans No. 680051 (SNABO), the Academy of Finland through its  Centres  of  Excellence  Program  (project  nos  312300, 312059, and 312294) and grants (Nos.  314447, 314449,  276528, 305237, 308161 and 314302), the Vilho, Yrjö and Kalle Väisälä Foundation,  the Technology Industries of Finland Centennial Foundation, the Jane and Aatos Erkko Foundation, and the Finnish Cultural Foundation.

\section*{Author contributions}
R.K. participated in the measurements and data analysis. V. V. integrated the JPA to the measurement setup, and participated to the measurements and data analysis. J.G., K.Y.T., and R.L. developed the fabrication process for the detector sample.  J.G. and R.L. designed and fabricated the detector sample. J.G designed the measurement setup. A.M.G carried out preliminary measurements with J.G. All authors have contributed to preparing the manuscript, although most of the work was carried out by R.K., J.G, V.V, and M.M. V.V., S.S., L.G., J.L., M.P., J.H., and O.-P.S. provided the JPA. M.M. supervised the work.

\section*{Competing interests}
The authors declare no competing interests.

\end{document}